\documentclass[12pt,preprint]{aastex}
\begin{document}
\def\beq{\begin{equation}}
\def\eeq{\end{equation}}
\def\bey{\begin{eqnarray}}
\def\eey{\end{eqnarray}}
\def\pc{\, {\rm pc} }
\def\kpc{\, {\rm kpc} }
\def\msun{M_\odot}
\def\sun{\odot}
\def\lsim{\mathrel{\raise.3ex\hbox{$<$\kern-.75em\lower1ex\hbox{$\sim$}}}}
\def\gsim{\mathrel{\raise.3ex\hbox{$  $\kern-.75em\lower1ex\hbox{$\sim$}}}}
\def\Msun{M_\odot}
\def\Lsun{L_\odot}
\def\lsun{L_\odot}
\def\kms{\, {\rm km \, s}^{-1} }
\def\eV{\, {\rm eV} }
\def\keV{\, {\rm keV} }
\def\dis{{\rm dis}}
\def\grad{{\bf \nabla}}
\def\HSZ{{\it HSZ}}

\title{Reinterpreting MOND: coupling of Einsteinian gravity and spin of cosmic neutrinos?}
\author{HongSheng Zhao\\Scottish University Physics Alliance, University of St Andrews, KY16 9SS, UK}
\begin{abstract}
Several rare coincidences of scales in standard particle physics 
are needed to explain (i) why neutrinos have mass, 
(ii) why the negative pressure of the cosmological dark energy (DE) coincides
with the positive pressure of random motion of dark matter (DM) in bright galaxies, (iii) why Dark Matter in galaxies seems to have a finite phase-space density, and to follow the Tully-Fisher-Milgrom relation of galaxy rotation curves.  The old idea of self-interacting DM is given a new spin: 
we propose that the neutrino spin-gravity coupling could lead to a cosmic neutrino dark fluid with a 
an internal energy density varying as function of the local acceleration of the neutrino fluid with respect the CMB background.    
We link the Tully-Fisher-Milgrom relation of spiral galaxies (or MOND) with the relativistic pressure of the neutrino dark fluid without modifying Einsteinian gravity. 
\end{abstract}

\maketitle


It is perfectly conceivable to have a universe with 
zero cosmological constant and zero mass of neutrinos, and in fact it is the most logical construction of the universe to preserve the most appealing symmetries of particle physics.  Yet symmetry is slightly broken in our universe for mysterious reasons: we have a tiny cosmological constant of space-time and a small mass of neutrinos.  The amplitude of the two 
are also comparable in terms of the energy density, $\rho_\Lambda \sim (0.001eV)^4$, 
and $\rho_\nu = (0.0001eV)^3 \times 1eV$ for neutrinos of eV range.  Mass-varying Neutrinos (MaVaNs)  has been proposed, assuming that these two mysteries could be related (Fardone et al. 2004, Kaplan et al. 2004), e.g.,  the mass of the neutrino could be due to coupling to a Higgs-like scalar field, whose energy density gives the cosmological constant effect.   Such a coupling leads to a time-dependent effective mass of the neutrino, which is unusual but allowed by physics.    

Given that ordinary neutrinos and sterile neutrinos are hot and cold dark matter candidates respectively, the natural question is to ask if dark matter could also be be unified by the same fields.  Unification of dark matter and dark energy have been an inspiration for many theories, for example, the Chapligin gas and the k-essence.   Many scalar field theories can yield a equation of state parameter $w$ for the dark stress between 0 and -1.   The naive expectation is that in the absence of interactions, neutrinos have large free-streaming length, hence cannot drive structure formation on the small scale.  Furthermore the oscillation experiments favor the neutrino mass around 0.05eV, far from dominating the baryons.  However, self-interaction in the neutrino sector could modify neutrino's properties.   It is known that the MaVaNs
are unstable to perturbations (Afshordi et al 2006), and in fact behave very much like Cold Dark Matter, hence can collapse into neutrino lumps at least on the galaxy cluster scale (Mota et al. 2008).

A harder question is whether the DM effects of the MaVaNs-like self-interacting field could also reproduce the Tully-Fisher relation.  
Self-interacting dark matter (Spergel \& Steinhardt 1999) and its variations have been studied extensively in the past.  A fixed cross-section per unit mass could help on some scales, but appears to be problematic on other scales.  This suggests a gradual tapering the cross-section might be helpful.  
Recently it has been proposed that the DM and DE effects could 
be unified by a Dark Fluid described by a single vector field (Zhao \& Li 2008).    The key property of this Dark Fluid is a non-linear pressure $P$ depending on the gravitational stress $|\nabla \Phi|^2/(8 \pi G)$; this was achieved by taking gradient of a time-like unit vector field.  Here we propose to link the neutrinos with the Dark Fluid of non-linear pressure.  The challenge is to give neutrinos the appropriate amount of self-interaction on various scales, perhaps making its effective mass increasing on larger and larger structure, and so that its property is between that of collisional fluid and ballistic particles free-streaming on geodesics.  


Neutrinos decouple from electrons or matter at $z_{dec} \sim 10^{10}$ with a thermal momentum distribution function 
\beq
f(p) = (2\pi \hbar)^{-3}  \left[ \exp( y ) + 1 \right]^{-1}, \qquad y \equiv {p \over  p(z) }, 
\qquad p(z) \sim 0.00018 (1+z){\rm eV/c}
\eeq
After decoupling, the expansion of the universe makes this distribution dilate with a characteristic momentum $p(z)$ or "temperature" $\sim 1.95(1+z)$Kelvin, or deBroglie wavelength of $\sim 0.1 (1+z)^{-1} $cm.  

It is commonly accepted that observed three flavors of neutrinos and their anti-flavors are the same Majorana particle with mixed states of flavors, helicities and masses.  This means the particle can start its life from the decoupling of Big Bang as a burst of, e.g., ultra-relativisitic neutrinos of pure electron flavor of momentum $p(z_{dec}) \sim 10^6$eV, it then is decomposed into three mass states of mass $0 \le m_1< m_2< m_3$, and changes flavors back and forth among electron, muon or tau types.  The expansion of the universe cools the neutrinos, such that at present day it can have  speeds $p(0)/m_1 \sim p(0)/m_2 \sim  p(0)/m_3 \sim 30 {m_3 \over 2{\rm eV}}$km/s, where $p(0) \sim 0.00018$eV/c, and we assume the three masses are comparable.  The slow neutrinos will be accelerated by the growing localized gravitational potential wells, and likely become bound.  Observers orbiting inside a galaxy cluster or even inside the solar system can sometimes overtake the slow cosmic neutrinos, and identify them instantaneously as  a flux of anti-neutrinos of reversed helicity.  
The fraction of bound vs unbound neutrinos or anti-neutrinos is a measurement of the potential well $\Phi(t,x)$, hence the gradient of the bound neutrino number density $\partial_x n(t,x)$ is a measurement of gravity $\partial_x  \Phi(t,x)$. 

It is interesting if $m_1=0$, or $m_1 \ll m_2 \sim m_3$, then a cosmic electron neutrino's wave function would be split into two parts, an unbound ultra-relativistic mass eigenstate $m_1 \sim 0$ plus a bound mixture of neutrinos and anti-neutrino mass eigenstates $m_2$ and $m_3$.  
So electron neutrinos are no longer localized particles, but are a probability distribution over the scale of the neutrino horizon, which is only slightly smaller than the light horizon $\sim cH_0^{-1}$ by a factor $\sim {v \over c} \sim {p(z) \over \sqrt{m_1^2 c^2 + p(z)^2 } }$.  While the detailed quantum treatment of the problem is beyond the scope of this paper, pressure-like long-range interaction has also been discussed by Pfenniger et al. (2007).
The bound neutrinos in galaxies will experience a pressure-like interaction, mediated by a background of ultra-relativistic neutrino mass eigenstate $m_1$.  This might create the impression of a modified gravity or modified inertia.  

At a classical level, neutrinos might be treated as a fluid with certain pressure, hence the flow of neutrinos is neither on null geodesic as photons nor massive geodesics  ${D \over D\tau} u_\alpha= u^\beta \partial_\beta u_\alpha -  \Gamma^{\gamma}_{\beta \alpha} u_\gamma u^\beta = 0$ like the CDM particles.  Rather the flow is subject to a pressure gradient due to self-interaction, so the flow speed's Lagrangian co-variant derivative ${D \over D\tau} u_\alpha \sim - {s(n)^2 \over n} \partial_\alpha n \sim s(n)^2 \partial_\alpha  \ln[(1+z) (1+2 \Phi(t,x))]$, where $s(n)$ is the fluid sound speed, which is likely a function of neutrino energy density $n$, whose inhomogeneity tracks the uneven gravitational potential $\Phi(t,x)$.   

What gives mass to active neutrinos is an open question, and it is not even clear whether the mass is constant or not.  Interaction with Higgs-like field in other sectors or perhaps self-interactions in the neutrino sector (perhaps with sterile neutrinos) give an effective mass $m(z,x)$ to the chirially left-handed observed neutrinos.  
This argues that the active neutrinos form self-collisional fluid with less free-streaming.
Since the exact mechanism of interactions are unknown, one could assume most generally that the effective mass could depend on environment.  Here we consider models in which the internal energy of neutrinos depend on the number density $n$ and its time evolution and gradient $\nabla n$.  

Unlike MaVaNs, here we assume that neutrinos have zero mass in the absence of perturbations, and the effects of mass appear only in collapsed systems, like galaxies and the solar system.  
We further assume that the neutrino pressure $P$ is some function of a scalar field and  
its temporal and spatial gradient.  If the scalar field is the 
temperature $T$ or number density $n$ of the massless neutrinos, then 
$P = P(T,  T^{-1} \dot{T},  T^{-1} \nabla T)$.
The temperature $T(x,t)$ of massless neutrinos varies in essentially the same way (up to a constant factor 0.75) as the cosmic photon background, namely, 
$T/T_0 \sim  (n/ n_0)^{1/3}  \sim \left( 1+ {2 \Phi(x,t) \over c^2} \right) / R(t) $
where presently $T \sim T_0=2$ Kelvin on average, and drops with the cosmic time as $1/R(t)$ and is slightly inhomogeneous due to gravitational potential $\Phi(t,x)$; here $R(t)$ is the scale factor of the universe, and $R=1$ presently.    We can replace $ \dot{n}/(3n) \sim  \dot{T}/T \rightarrow  \dot{R}/R + \dot{\Phi}/c^2$ and $n^{-1} \nabla n/3 = T^{-1} \nabla T \rightarrow \nabla \Phi/c^2$, 
and assume no preferred spatial directions.  The equation of state becomes $P = P(n, \dot{n}/n, |\nabla \Phi|)$, which depends on the density and its rate of change and spatial gradient (the local gravity $\nabla \Phi|$), e.g., $P = n^{\gamma} f(|\nabla \Phi|)$.  The original MaVaNs corresponds to a special case $f=cst$, and $\gamma \le 0$, so that the neutrinos can be modeled as a nearly polytropic fluid with an imaginary sound speed $P \propto n^\gamma$ in an expanding universe.  For our purpose on galaxies, we neglect cosmic expansion for simplicity, and consider galaxies in a static universe with a Minkowski background $R(t)=1$, and $n \sim n_0 R^{-3} = n_0 \sim 200{\rm cm}^{-3}$, which can be treated as a constant.   

To be specific we define $\beta$ to be the sound speed the relativistic neutrino fluid in units of the speed
of light $c$, and we adopt a variable sound speed such that  
\beq\label{pres}
{1 \over \beta}  =  1+ {A \over  |\nabla \Phi|  + B}, \quad {A^2 \over 8\pi G} \equiv  n E_0,
\eeq
where we set $B=0$ so that the dimensionless sound speed $\beta=0$ for the uniform universe\footnote{MaVaNs corresponds to an imaginary constant $B = \sqrt{-\Lambda}$.}, and
$E_0$ is certain fixed low-energy threshold in the neutrino sector respectively, whose physics is not specified here.  
The quantity $1/\beta$ carries the meaning of the ratio of gravitational wave speed $c$ vs the sound speed $c \beta$ of the neutrino fluid, or the "refraction index" of the neutrino fluid, which is assumed to be function of the ratio of the gravitational stress 
\beq
P_g = {|\nabla \Phi|^2 \over 8\pi G}
\eeq 
vs a characteristic stress $n E_0 \equiv {A^2 \over 8\pi G}$ of the neutrino fluid.  One can motivate such a correction by saying quantum effects must be considered 
to model interactions of the metric with a spin 1/2 neutrino fluid in a box of deBroglie wavelength  
$1/n \sim 1/n_0 \sim 5  {\rm mm}^3$; 
when the gravitational energy inside this box is just below or above certain threshold $E_0$, 
one might expect that anti-particles can disappear or appear (reminiscent of the famous Klein paradox), 
hence can change the  energy density of neutrinos and anti-neutrinos, hence the sound speed of the neutrino fluid.  Interpreting in the picture of the the dielectric analogy of MOND (Blanchet \& Le Tiec 2008, Famaey, Gentile, Bruneton, Zhao 2007), one would say that the neutrino medium is ``polarized" by a variable amount due to a varying gravitational stress.     
The adopted refraction index is such that the perturbations in the stress of the neutrino fluid propagate as the speed of light (as in vacuum) in strong gravity, but much slower in weak gravity;  no propagation or pressure in Minkowski space, where $\nabla \Phi=0$.  In the solar system, where $\nabla \Phi$ is much bigger than $A$, cosmic neutrinos behave relativistically.

We consider a classical Newtonian theory where one minimizes the action $S$ given by
\bey
S &=& \int dt d (xR)^3 \left( \rho_b \Phi  + P_g -  P \right), \qquad P=(n E_0) \beta^2, 
\eey
where the relativistic pressure of neutrino fluid is included by a term $n E_0  \beta^2$, and 
$P_g + \rho_b \Phi$ 
is the Lagrangian of gravitational field and the baryons of density field $\rho_b$, which couples to gravitational potential $\Phi$.  Here we use the non-covariant formulation of gravity as an external field, which is the weak perturbation limit of General Relativity.  



The total Lagrangian now resembles the classical MOND of Bekenstein-Milgrom (1984), 
\beq
L = {A^2 \over 8 \pi G} (y-F(y)) + \rho_b \Phi , 
\eeq
where 
\beq
F(y)  =  {y \over (\sqrt{y}+1)^2}, \qquad y  \equiv {|\nabla \Phi|^2  \over A^2}. 
\eeq

Apply the Lagrangian equation by varying the total Lagrangian with respect to $\Phi$, we obtain the  MOND-like Poisson equation
\beq
2 \nabla \left[ \mu \nabla \Phi \right] = 8 \pi G \rho_b,
\eeq
where
\beq
\mu = 1- dF/dy = 1- (1+\sqrt{y})^{-3}.
\eeq

It is interesting that a relativistic neutrino fluid with a non-trivial pressure can give the physics of MOND.
To match with MOND acceleration, we can set 
\beq
A =3 a_0 = 3.6 \times 10^{-10}{\rm m/s}^2.
\eeq
The resulting $\mu$ function here matches that of Zhao (2007), which are shown to be compatible with solar system data and spiral galaxy rotation curves, especially the Tully-Fisher-Milgrom relations. 
The MOND acceleration scale would corresponds to a mass or energy scale in the neutrino sector
\beq
E_0 = (3a_0)^2/(8 \pi G n_0) \sim 0.1 { 3 H_0^2 \over 8 \pi G n_0} \sim 2{\rm eV}.
\eeq  
Note that any eV range neutrinos with a normal non-relativistic pressure is  insignificant to galaxy potential.  The  non-linear coupling of the neutrino pressure to the metric through some unspecified quantum effects is the key.  In our picture neutrinos are not localized, this can create the effects of a ubiquitous dark matter fluid or modified gravity. 


Let's compare with previous models.  
The neutrino model here is different from both traditional hot dark matter model and from MaVaNs, since we do not require a rest mass for neutrinos in the limit of a uniform universe.  
We could relax the assumption that neutrinos are massless in a homogeneous isotropic universe,
hence create a model somewhere in between a hot dark matter model and pure MOND (Angus et al. 2007).    A non-zero mass of neutrinos in the absence of perturbations 
would contribute to the Dark Matter budget in the Hubble expansion.  
Letting the mass change with the neutrino number density would create a dark pressure as in MaVaNs.  

There are also important differences of our neutrino model and the classical MOND Modified Gravity formulation of a similar effect.  Unlike in MOND where $a_0$ is fixed at about $ \sim H_0/6$, our acceleration scale $A$ in principle varies with redshift, as $A \propto \sqrt{n} \propto (1+z)^{3/2}$.  The fact that the neutrino sector has a mass scale as required by the neutrino oscillation experiments, could be the physics of the MOND coincidence puzzle, i.e, why $a_0$ "happens" to be comparable to present $H_0$, i.e.,  any mass scale in the neutrino sector will set a common scale on both dark energy and (galactic) dark matter, so $A$ will track the Hubble expansion.
A non-zero mass would allow neutrinos to condense into galaxy clusters, the neutrino number density $n(x,t)$, hence $A(x,t)$ would be spatially dependent.  The MOND effects would be made bigger in galaxy clusters, plus the fact that neutrinos add to the total mass of the system.  This could be a welcome direction to mend the problems of MOND in galaxy clusters.   This idea is also different from the idea of adding plain massive neutrinos in galaxy clusters (Sanders 2003, Angus et al. 2007).  However, a rigorous co-variant model of the evolution of the neutrino number density involves solving the Boltzmann equation for perturbations in a cosmological background, which deserves  further study.
  
One advantage of our neutrino model here is that it recovers the Tully-Fisher relation, hence 
it is immune from the lasting problems of CDM, i.e., over-prediction of dark matter density in galaxy centers and over-prediction of dark matter in dwarf galaxies (Gnedin \& Zhao 2002 and references therein).   While centres of CDM halos can have arbitrary density and pressure, the neutrino fluid (cf. eq.~\ref{pres}) has a limited range of its pressure $P = n E_0 \beta^2$.  There is a maximum pressure difference of 
\beq
|P_1 - P_2  | =  n E_0 \left[ \left({ B+ |\nabla \Phi|  \over A+ B + |\nabla \Phi| }\right)^2-\left({B \over A+B} \right)^2 \right] \le n E_0 \left[1 - \left({B \over A+B}\right)^2\right] \le n E_0 
\eeq
between the center of a dense galaxy where $|\nabla \Phi| \gg A = \sqrt{8\pi G n E_0}$ 
and the cosmic average where $|\nabla \Phi |\sim 0$.   
This pressure of neutrino appears a key ingredient missing in standard Dark Matter models.  
The present amplitude of this pressure $n_0 E_0 \sim 0.1 $ times the critical density, hence is comparable to the Dark Energy observed.  

This pressure can be understood as collisional cross-section of self-interacting dark matter. Incidentally
collisonal dark matter is also related to the tight correlation of black hole mass and velocity dispersion relation of galaxies, which can be understood by a balance of hydrostatic pressure (Xu et al. 2007) or 
the constant refilling of the loss cone by DM collisions (Zhao, Haenalt, Rees 2002), or 
the Bondi accretion of fluid dark matter up the mean free path (Ostriker 2000).  
In short a non-conventional neutrino could relate  
several important puzzles ranging from neutrino oscillation in the solar system to galactic astronomy and cosmology.    

{}


\begin{thebibliography}{0}
\bibitem[]{} Afshordi N., Zaldarriaga M., Kohri K., 2005, PRD, 72, 065024, arXiv:astro-ph/0506663
\bibitem[]{543} Angus G.W., Shan H, Zhao H., Famaey B., 2007, ApJ, 654, L13 
\bibitem[]{Bek_01} Bekenstein J., 2004, Phys. Rev. D., 70, 3509
\bibitem[]{546} Bekenstein J., \& Milgrom M. (1984), ApJ, 286, 7 (BM84)
\bibitem[]{} Blanchet L., Le Tiec, A., arXiv0804.3518B
\bibitem[]{} Mota D., Pettorino V., Robbers G., Wetterich C., 2008, PhLB, 663, 160
\bibitem[]{} Famaey B., Gentile G, Bruneton J.P., Zhao H., 2007, PRD, 75, 3002 
\bibitem[]{} Fardone R., Nelson A.E., Weiner N. 2004, JCAP, 0410, 005 arXiv: astro-ph/0309800
\bibitem[]{} Kaplan D.B., Nelson A.E., Weiner N. 2004, PRL, 93, 091801 arXiv: hep-ph/0401099
\bibitem[]{}  2004, JCAP, 0410, 005 arXiv: astro-ph/0309800
\bibitem[]{553} Gentile G., Famaey B., Tiret O., Combes F., Kroupa P., Zhao H., 2007, A\&A Lett., 472, L25
\bibitem[]{510} Gnedin O. \& Zhao H., 2002, MNRAS, 333, 299
\bibitem[]{560} McGaugh S. 2005, Phys. Rev. Letters, 95, 1302
\bibitem[]{} Ostriker J.P., 2000, PRL, 84, 5258
\bibitem[]{San} Sanders R.H., 2003, MNRAS, 342, 901
\bibitem[]{} Spergel D. \& Steinhardt P. 1999, PRL , arXiv:astro-ph/9909386
\bibitem[]{526} Xu B.X, Wu X.B., Zhao H., 2007, ApJ, 664, 198
\bibitem[]{} Zhao H.S., Li B. 2008, ApJ submitted, arXiv0804.1588
\end{thebibliography}
\end{document}